\documentclass[12pt]{iopart}
\usepackage{iopams}
\usepackage{epsfig}

\begin{document}

\newcommand\changes{\marginpar {CHANGES} }
\newcommand\boh{\marginpar {?} }
\newcommand{\sq}{\sqrt{3}}
\newcommand{\di}[1]{\frac{#1}{\sqrt{3}}}
\topmargin 0pt \oddsidemargin 5mm \headheight 0pt \headsep 0pt
\topskip 9mm
\newcommand{\X}{\bar{X}}
\newcommand{\MM}{{\mathbb M}^{(1,3)}}
\newcommand{\MMM}{{\mathbb M}^{(1,4)}}
\newcommand{\dS}{{dS_4}}
\newcommand{\eq}{\begin{equation}}
\newcommand{\feq}{\end{equation}}
\newcommand{\eqn}{\begin{eqnarray}}
\newcommand{\feqn}{\end{eqnarray}}
\newcommand{\arr}{\begin{eqnarray*}}
\newcommand{\farr}{\end{eqnarray*}}
\newcommand{\bea}{\begin{array}}
\newcommand{\ea}{\end{array}}
\newcommand{\de}{\partial}
\newcommand{\tildem}{\mu}
\font\mybb=msbm10 at 12pt
\def\bb#1{\hbox{\mybb#1}}
\def\bR {\bb{R}}
\newcommand{\cHH}{{dS}}
\newcommand{\diag}{\rm diag}
\newcommand{\D}{d}
\newcommand{\A}{\mu}
\newcommand{\B}{\nu}
\newcommand{\oo}{X_0}
\newcommand{\YY}{\bar Y}
\newcommand{\XX}{\bar X}

\title{Conservation laws and scattering for de Sitter classical particles}

\author{S Cacciatori$^1$\footnote{E-mail address: sergio.cacciatori@uninsubria.it},
V Gorini$^1$\footnote{E-mail address: vittorio.gorini@uninsubria.it}, A Kamenshchik$^{2,3}$\footnote{E-mail address: alexander.kamenshchik@bo.infn.it}
and  U Moschella$^1$\footnote{E-mail address: ugo.moschella@uninsubria.it}}

\address{$^1$Dipartimento di Fisica e Matematica, Universit\`a dell'Insubria\\ Via Valleggio 11, 22100 Como, Italy and INFN, Sez. di Milano, Italy}

\address{$^2$Dipartimento di Fisica and INFN, via Irnerio 46, 40126 Bologna, Italy}

\address{$^3$ L D Landau Institute for Theoretical Physics of the RAS, Kosygin street 2, 119334 Moscow, Russia}

\begin{abstract}
Starting from an intrinsic geometric characterization of de Sitter
timelike and lightlike geodesics we give a new description of the
conserved quantities associated with classical free particles on the
de Sitter manifold. These quantities allow for a natural discussion
of classical pointlike scattering and decay processes. We also
provide an intrinsic definition of energy of a classical de Sitter
particle and discuss its different expressions in various local
coordinate systems and their relations with earlier definitions
found in the literature.
\end{abstract}

\maketitle

\section{Introduction}
Since the first pioneering observations of the luminosity-red shift
relation of distant type Ia supernovae
\cite{Perlmutter:1998np,Riess:1998cb,Riess:2001gk,Spergel:2003cb} it
is by now accepted as an established fact that the expansion of the
universe is accelerated. This circumstance could be interpreted by
saying that there exists some kind of agent, dubbed dark energy,
which exerts an overall repulsive effect on ordinary matter (both
visible and dark). This repulsion has long since overcome the mutual
attraction of the various parts of the latter, thereby being
responsible for the present accelerated expansion. The nature of
dark energy is to date entirely mysterious. The only facts we know
with reasonable certainty are that dark energy contributes today in
the amount of about 73 \% (the exact figure depending on the
cosmological model adopted) to the total energy content of the
universe, and that its spatial distribution is compatible with
perfect uniformity.

The simplest possible explanation for dark energy which can be put
forward is to assume that it is just a universal constant, the so
called cosmological constant, denoted $\Lambda$. If we espouse this
point of view,  this would mean that the background arena for all
natural phenomena, once all physical matter-energy has been ideally
removed, is not the familiar flat Minkowski spacetime $\MM$.
Instead, that it consists of the maximally symmetric de Sitter
spacetime $\dS$ whose radius $R$ is related to $\Lambda$ by the
equation $R=\sqrt{3/\Lambda}$. The actual value of $\Lambda$ is
extremely small in astrophysical and also in galactic terms
($\Lambda \simeq 10^{-56} \makebox{cm}^{-2}$), so that cosmic
expansion has no significant effect say on the structure of a
typical galaxy, such structure being essentially controlled by the
material (in all its forms) composing the galaxy itself, by the
mutual gravitational attraction of the galaxy's parts and by the
galaxy's angular momentum. On the other hand, $\Lambda$ has an
essential effect on the distribution of matter on large cosmic
scales, such as on the structure of the cobweb pattern of filaments
and voids characterizing the arrangements of galaxies and galaxy
clusters in the universe.

It is not our purpose here to deal with the by now longstanding
problem of the nature of dark energy and of why the dark energy
content of the universe is, at the present epoch, comparable with
the universe's ordinary matter content. See e.g. the reviews
\cite{Sahni:1999gb, Padmanabhan:2002ji,Peebles:2002gy}. Instead, we
adhere to the simple working hypothesis that the cosmological
constant is a true universal constant, just like such are the speed
of light and Planck constant, say.

In the approximation in which the effects of gravitation on the
geometry of spacetime can, at least locally, be neglected, the
presence of a cosmological constant would naturally lead to the
problem of the formulation of the theory of special relativity in
presence of a universal residual constant background curvature,
namely of a de Sitter relativity in place of the customary flat
Minkowski one. Then, the symmetry group of the theory would be the
de Sitter group $SO(1,4)$ (the Lorentz group in five dimensions)
instead of the Poincar\'e group, which is the contraction \cite{IW}
of the latter arising in the limit $\Lambda \to 0$.  A considerable
amount of work has already been performed in this direction, both in
the classical
\cite{Abbott:1981ff,Aldrovandi:2004hd,Aldrovandi:2006vr,Guo:2003qm,Guo:2004pj,Guo:2006pa,KowalskiGlikman:2003we,Gursey:1964}
as well as in the quantum domain
\cite{Birrell:1982ix,Bros:1995js,Bros:1998ik,Bros:2006gs,Ugo:2005,Ugo:2007}.
However, to our knowledge, at an elementary level a systematic
treatment of particle kinematics and dynamics in de Sitter spacetime
still lacks.

In this paper we contribute to  fill this gap by providing a
description of  the free motion of classical particles and of
particle collisions in terms of an intrinsic characterization of the
associated conservation laws.  We adhere, of course, to the geodesic
hypothesis \cite{Synge}: the worldline of a free particle is a
geodesic in spacetime. In the de Sitter universe timelike and
lightlike geodesics can be fully and economically characterized by
using the closest analogue to Minkowski momentum space that is
available: this is the lightcone of the five-dimensional Minkowski
space ${\mathbb M}^{(1,4)}$ in which the de Sitter universe can be
represented as an embedded four-dimensional one-sheeted hyperboloid.
The relevant conserved quantities associated with free motion can
themselves be expressed in terms of the same lightlike five-vectors,
as we do here. Then, it turns out that, in a given particle
collision, the conservation of energy and momentum of ingoing and
outgoing particles at the collision point can be expressed in terms
of the corresponding one particle conserved quantities before and
after the collision.

The structure of the paper is as follows. In Section
\ref{sec:conservation} we first recall the expression of the
generators of the de Sitter symmetry group  in terms of the
flat coordinates of the five-dimensional ambient Minkowski
space. Then, by using Noether theorem applied to the invariant
action of a free massive particle we derive the set of the
associated conserved quantities $K$. Of such conserved
quantities we give two different intrinsic characterizations.
One in terms of the two lightlike vectors $\xi$ and $\eta$ of
$\MMM$ that uniquely identify  the given timelike geodesic. The
other one in terms of either one of such vectors and of a given
point of the geodesic. These characterizations are independent
of the choice of any particular coordinate patch on the de
Sitter manifold $\dS$. We also find the corresponding formulae
for lightlike geodesics.

In Section \ref{sec:collisions} we describe particle collisions
and decays in terms of the conserved quantities introduced
earlier. Precisely, we re-express the conservation of the total
energy-momentum at the point of a collision as a conservation
law for the total invariants $K$. In particular, the
conservation equations can be given a perspicuous expression
which involves explicitly the collision point. The conservation
of the invariants $K$ allows us to relate the values of the
energy and momentum at the point of collision to their values
at any observation point. We do this by providing an explicit
formula, valid both for massive  and massless particles, which
indeed relates the energy-momentum vector at two arbitrary
points on the geodesic. In particular, this formula applied to
photons yields the well-known frequency redshift relation.

Section \ref{sec:energy} is devoted to the definition of the
energy of a free particle, both massive and massless, by
comparison of the corresponding geodesic to the reference
geodesic associated with a {localized} observer. This
definition is itself intrinsic and  does not make reference to
any particular coordinate patch. However, we also give the
explicit expression of the energy in terms of some specific
coordinate choices on the de Sitter manifold: flat, spherical
and static coordinates, the first two having cosmological
significance, the third chosen for its relevance in black hole
physics. To our knowledge, this definition of the energy of a
de Sitter particle first appeared in \cite{Gursey:1964}; it was
introduced there in yet another set of coordinates, the
stereographic ones. The value and the novelty of our approach
resides in the fact that we have given to the definition of
energy a coordinate independent meaning.

Section \ref{sec:momentum} is devoted to a possible definition of
particle momentum. While the definition of energy
 is intrinsic, being purely related to a reference geodesic, any possible definition of
momentum is unavoidably linked to a choice of a  coordinate system.
Nevertheless, we examine  reasonable expressions of momenta
corresponding to different choices of coordinates. The consistency
of these definitions is set in evidence by the fact that in the flat
Minkowski limit all these choices converge to the correct flat
momenta expression.

We end with several concluding remarks.

\section{Conservation laws for de Sitter motion.}\label{sec:conservation}
In what  follows we will present our results by making
reference to the (physical) four-dimensional de Sitter
spacetime. However, as it will be evident from the discussion,
our formulae are completely general and valid in any dimension
just by replacing 4 by $d$ (and $5$ by $d+1$).

The $4$-dimensional de Sitter spacetime $\dS$ can be realized
as the  one-sheeted hyperboloid with equation
\begin{equation}
\dS = \{X\in \MMM , \   X^2= X\cdot X=\eta_{AB}X^A X^B = -R^2\}
\end{equation}
embedded in  the $5$-dimensional Minkowski spacetime $\MMM$
where a Lorentzian coordinate system has been chosen: $X =
X^A\epsilon_A$ and whose metric is given by
$\eta_{AB}=\diag\{1,-1,-1,-1,-1\}$ in any Lorentzian frame. The
geometry of the de Sitter spacetime is induced by restriction
of the metric of the ambient spacetime to the manifold:
\begin{equation}
ds^2= \left.(\eta_{AB} dX^A dX^B)\right|_{\dS}.
\end{equation}
This is the maximally symmetric solution of the cosmological
Einstein equations in vacuo provided that $R =
\sqrt{3/\Lambda}$, with $\Lambda>0$. The corresponding isometry
group (the relativity group of $dS_4$) is $SO(1,4)$, i.e. the Lorentz group of the ambient
spacetime $\MMM$ which is generated by the following ten Killing vector
fields\footnote{The restriction of these operators to the de
Sitter manifold  is well-defined. This can be shown by
introducing the projection operator $h$ and the tangential
derivative $D$ as follows:
\begin{equation*}\label{}
h^{ A  B } = \eta^{ A  B } + \frac{X^{A} X^{B}}{R^2}, \ \ \ \ D^{A } =
h^{ A  B }{ \partial}_{ B }= { \partial}^{ A } +  \, \frac {X^{A}}{R^2}\, X\cdot \partial.
\end{equation*}
It follows that
\begin{equation*}
L_{AB}  = X_A \partial_B - X_B\partial_A = X_A D_B - X_B D_A \ .
\end{equation*}}
\begin{eqnarray} L_{AB}=\left.\left( X_A \frac {\de}{\de X^B} -X_B \frac
{\de}{\de X^A} \right)\right|_{\dS}.
\end{eqnarray}
Since the group acts transitively on the manifold $dS_4$ it is
useful to select a reference point (the origin) in $\dS$ as
follows:
\begin{equation}
\oo = (0,0,0,0,R).
\label{origin}
\end{equation}

Now consider a classical  massive particle on the de Sitter
universe. The usual action for geodesical (free) motion can be
written by using the coordinates of the ambient five-dimensional
spacetime as follows:
\begin{eqnarray} S=-mc\int \left[(V^2)^{\frac{1}{2}} + a (X^2+R^2)\right]\ d\lambda \, ;\label{freeaction}
\end{eqnarray}
here $\lambda \to X(\lambda)$ is a parameterized timelike curve
subject to the constraint $X^2(\lambda) = -R^2$  as enforced by the
Lagrange multiplier $a$; $V^A(\lambda)={dX^A}/{d\lambda}$ is the
corresponding velocity. $V^A(\lambda)$ is tangent to the curve
$X(\lambda)$ and therefore orthogonal to the vector $X(\lambda)$ (in
the ambient space sense). The condition of tangentiality $X\cdot
V=0$ has to be imposed also on the initial conditions when solving
the equations of motion. Consider now the generic infinitesimal
isometry of $dS_4$
\begin{eqnarray} X_A \longmapsto X_A +\omega_{AB} X^B \ , \label{isometry}
\end{eqnarray} where $\omega_{AB}$ are antisymmetric
infinitesimal parameters. The action is invariant under (\ref{isometry}) and using Noether theorem we find ten
quantities that are conserved along the timelike geodesics:
\begin{eqnarray} K_{AB}&=&\frac {m(X_A V_B
-X_B V_A)}{R\sqrt { V^2}} =\frac mR (X_A W_B-X_B W_A) =\cr &=&\frac 1R
(X_A \Pi_B -X_B \Pi_A) ; \label{duesei}
\end{eqnarray}
 $ W^A ={dX^A}/{d\tau}$ is the Minkowskian five-velocity relative to
the proper time $d\tau = ds/c$ and $\Pi^A=m\frac {dX^A}{d\tau}$ the
corresponding Minkowskian five-momentum. Of these ten quantities
only six are independent. Indeed, in order to specify a geodesic
completely one must for example assign the proper initial
conditions, namely the initial point on $dS_4$ and the initial
velocity (at $\tau=0$, say). We note that the quantities
(\ref{duesei}) are of course defined also along particle
trajectories which are not geodesics. However, in this case not all
of them (if any) will be constants of the motion.

We now derive two alternative intrinsic characterizations of the
conserved quantities $K_{AB}$ which are independent on the choice of
any particular coordinate patch on $\dS$. We do this by exploiting
an elementary way to describe the de Sitter timelike geodesics: in
complete analogy with the great circles of a sphere that are
constructed by intersecting the sphere with planes containing its
center, the de Sitter timelike geodesics can be obtained  as
intersections between $\dS$ and two-planes containing the origin of
$\MMM$ and having three independent spacelike normals. Each such
two-plane also intersects the forward lightcone in $\MMM $ (the
asymptotic cone)
\begin{eqnarray}
C^+ = \{X\in \MMM, \ X^2=0, \ X^0>0\}\ ,
\end{eqnarray}
along two of its generatrices. Any two future directed null
vectors $\xi$ and $\eta$  lying on such generatrices (see
figure)  can be used to parameterize the corresponding geodesic
in terms of the proper time as follows \cite{Ugo:2007}:
\begin{equation}
X (\tau) = R \frac { \xi\, e^{\frac {c\tau}R}-\eta \,e^{-\frac
{c\tau}R}}{\sqrt {2\xi\cdot\eta }}\ .\label{parametrizzazione}
\end{equation}

\begin{figure}[h]
\begin{center}
\includegraphics[height=8cm]{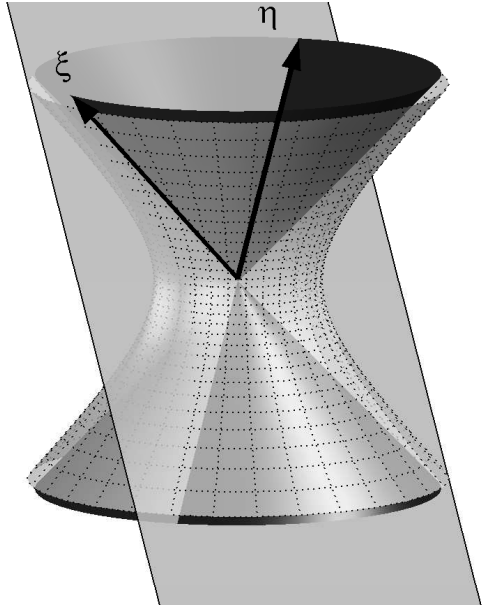}
\caption{\em Construction of a timelike geodesic of the de Sitter
manifold.
The asymptotic future lightcone of the ambient
spacetime; the vectors $\xi$,  $\eta$ belonging to $C^+$ play the role of
momentum directions.} \label{figgeo}
\end{center}
\end{figure}
Then, by inserting (\ref{parametrizzazione}) into (\ref{duesei}) we
find that the conserved quantities  have a very simple expression, homogeneous of degree zero,  in  the components of the vectors $\xi$ and $\eta$:
\begin{eqnarray} K_{AB}=mc \frac {\xi_A
\eta_B-\eta_A\xi_B}{\xi\cdot\eta} \ .\label{massconserved}
\end{eqnarray}
These numbers also coincide with the components  of the two-form\footnote{$\xi$ and $\eta$ denote here the covariant one-forms associated
to the null vectors; we use the same symbol for a
vector and its dual.}
\begin{eqnarray} K= K_{(\xi,\eta)} = mc \frac {\xi \wedge \eta}{ \xi\cdot\eta} \ , \label{210}\end{eqnarray}
in the  frame $\{\epsilon_A\}$ that has been chosen
in the ambient space. We normalize the dimensionless vectors $\xi$ and $\eta$ according with
\begin{eqnarray}
\xi \cdot \eta =\frac {2m^2}{k^2}\ , \label{211}
\end{eqnarray}
where $k$ is a constant with the dimensions of a mass whose value can be fixed according to the specific convenience.
With this normalization, formulas (\ref{parametrizzazione}), (\ref{massconserved}) and (\ref{210}) write respectively
\begin{eqnarray}
 X (\tau)&=&\frac {k R}{2m} \left( \xi e^{\frac {c\tau}R}-\eta e^{-\frac {c\tau}R}\right)\ , \label{212}\\
 K_{AB}&=&\frac {k^2 c}{2m} (\xi_A \eta_B-\xi_B \eta_A)\ , \label{213}\ \ \ \
 K_{(\xi,\eta)}=\frac {k^2 c}{2m} (\xi\wedge\eta) \ .
\label{214}
\end{eqnarray}
The replacements $\xi\longrightarrow \mu\xi$, $\eta\longrightarrow
\mu^{-1} \eta$ ($\mu>0$),
 do not alter (\ref{211}). As (\ref{212}) shows, they do
however shift the origin of the $\tau$ variable. Therefore, the
normalizations of $\xi$ and $\eta$ are fixed separately by equation
({\ref{211}) (in which a given choice has been made for the positive
constant $k$) and by selecting the point on the geodesic
corresponding to zero proper time. With these qualifications it
turns out that the pair $(\xi,\eta)$ depends on six independent
parameters. Then (\ref{213}) shows once more that only six of the
ten constants of  motion $K_{AB}$ are independent (in the appendix
we illustrate this fact with an explicit example).

Formula (\ref{210}) (or equivalently formula (\ref{214}))
provides our first intrinsic characterizations of the constants
$K_{AB}$. The second characterization that we display brings
about an arbitrary fixed point $X (\tau)$ on the geodesics.
Indeed, from (\ref{212}) one has the relation
\begin{eqnarray}
\eta=\xi -\frac {2m}{kR} \XX\ , \label{215}
\end{eqnarray}
where $\XX = X(0)$, which allows to rewrite the geodesic
(\ref{212}) in the alternative form
\begin{eqnarray} X(\tau) = \XX e^{-\frac {c\tau}R}+ \frac {kR\xi }{m} \sinh
\frac {c\tau}{R} . \label{1fondamentale} \end{eqnarray}
Inserting (\ref{215}) into (\ref{213}) and using
(\ref{1fondamentale}) gives \begin{eqnarray} K_{AB}=\frac {kc}R
(X_A(0)\xi_B-X_B(0)\xi_A)=\frac {kc}R e^{\frac
{c\tau}R}(X_A(\tau)\xi_B-X_B(\tau)\xi_A)\ . \label{kappa}
\end{eqnarray} As before, we can introduce the tensor
\begin{eqnarray} K=K_{\xi,X}=\frac {kc}R e^{\frac {c\tau}R}
(X(\tau)\wedge \xi)\ . \label{218} \end{eqnarray} The
normalization (\ref{211}) and Eq. (\ref{215}) imply
\begin{eqnarray} \xi\cdot \XX =-\frac {Rm}k \ . \label{imply}
\end{eqnarray} The tensor $K$, in its two alternative
expressions (\ref{210}) and (\ref{218}) will play an important
role in the following.

To perform the massless limit we set
\begin{equation}
m=k\epsilon,\ \ \tau =\frac \sigma{c} \epsilon
\label{massless-limit}
\end{equation}
and let $\epsilon \longrightarrow 0$ in Eqs. (\ref{1fondamentale})
and (\ref{imply}), thus obtaining the parametrization of a
lightlike geodesic:
\begin{eqnarray}
X (\sigma) =\XX +\xi \sigma \label{nullgeod}\ \ \ \
\makebox{ with }           \ \ \ \
\xi\cdot \XX=0\ \label{221}
\end{eqnarray}
where $\sigma$ is an affine parameter. Therefore, a lightlike geodesic is characterized by one lightlike vector which
is parallel to the geodesic and by the choice of an initial event that uniquely selects the particular geodesic among
the infinitely many pointing in that direction.
The conserved quantities are still given by formula (\ref{kappa})
\begin{eqnarray}
\tilde K_{AB}=\frac {kc}R (X_A(0)\xi_B-X_B(0)\xi_A)=\frac 1R (X_A \Pi_B -X_B \Pi_A) \ , \label{222}
\end{eqnarray}
where $\Pi^A= kc \frac {dX^A}{d\sigma}$ is the Minkowskian five-momentum of the zero mass particle.
There is of course no analogue of formulas (\ref{210}) and (\ref{214}) because $\xi$ and $\eta$ coincide in the massless limit.
An alternative standard way to arrive at formulas (\ref{222}) starts from rewriting the action for a massive particle
in the first order formalism:
\begin{eqnarray}
S[e,\gamma]=\frac k2 \int_\gamma \left[ \frac 1e
V^2 +a (X^2+R^2)\right] d\lambda+\frac 1{2k} m^2c^2 \int_\gamma e d\lambda \ , \label{action}
\end{eqnarray}
where $e$ is a function of $\lambda$ and $k$ is once more a constant with the dimensions of a mass. The equations of motion are
obtained by varying the action with respect to $e$ and to the
curve 
$\gamma$. The action for massless particles is
obtained by setting $m=0$ in Eq. (\ref{action}) and the
corresponding equations of motion are
\begin{eqnarray}
&& V^2 =0 \ , \ \ \
\frac d{d\lambda} \left( \frac 1e V^A \right)=0 \ .
\end{eqnarray}
Introducing an affine parameter $\sigma$ such that $d\sigma =c
e(\lambda) d\lambda$, the general solution is (\ref{nullgeod}). The
conserved quantities can be determined as before by means of
Noether theorem, giving (\ref{222}).
\subsection{Remarks on quantization.}
The setup that we have just described can also be employed to
provide a fresh look to de Sitter quantum mechanics and field
theory. Indeed, the variables on the cone in $\MMM$ that we have
been using to describe the geodesics can be employed to parameterize
the phase space pertinent to elementary systems. Then one can
invoke his favourite method, like geometric quantization
\cite{souriau,konstant} or the method of coadjoint orbits
\cite{kirillov} to obtain a quantum description of such elementary
systems. In doing this, a substantial difference will arise when
quantization deals with massless particles and the method will fail
to provide a de Sitter covariant theory. This problem has been known
for a long time and has fairly profound implications for
cosmological structure formation (since it gives rise to the scale
invariant spectrum of inflation \cite{harrison,zeldovich,mukhanov})
and for the dynamical restoration of spontaneously broken continuous
symmetries \cite{Ratra}. These effects are genuine quantum phenomena
and have no classical counterpart. We shall not investigate further
geometric quantization here and leave it for future work.

However, we can at least provide here an heuristic example of a connection between the classical and quantum counterparts
of a typical physical effect arising in the de Sitter manifold. For simplicity, we discuss this in two dimensional de
Sitter spacetime. Specifically, consider a ``rigid'' (here one dimensional) box of length $2L$ containing initially
a uniform distribution of a large number of identical point particles of mass $m$ which are all at rest relative to the endpoints
(walls) of the box. ``Rigid'' means here that we assume the internal forces holding the box together to prevent it from
participating unhindered to the de Sitter expansion, so that a local geodesic observer $O$ comoving with the box sees that the spatial extension
of the latter does not change in time. In other words, the walls of the box are not receding away during the cosmic expansion.
In particular, if we assume the observer $O$ to sit at the midpoint of the box, the worldlines of the endpoints of the box
will not be geodesics and the box itself will shrink compared to the comoving spatial coordinates. As to ``initially at rest''
it means that, at a given initial time, all the particles inside the box are assumed to move with zero velocity in the comoving frame.
Then, because of the expansion of the universe, the observer $O$ will see the particles move away from each others and eventually
start hitting the walls of the box, bounce back and collide with each other. By virialization, the result is that (after a long time and
in a nonrelativistic framework) they will reach a thermodynamic equilibrium at a temperature $T_c=H^2 L^2 m/2.26k$, where $H$ is
the Hubble constant and $k$ the Boltzmann constant (see appendix B).

Now consider a quantum scalar field in $dS_2$, which we assume to be in its ground state (the de Sitter vacuum \cite{Birrell:1982ix,bunch}).
Due to the interaction with the spacetime curvature the vacuum fluctuations generate real particles with a thermal spectrum at temperature
$T_q=\hbar c/2\pi Rk$ (\cite{gibbons,Bros:1998ik}). It seems reasonable to assume that the lightest allowed mass (the $dS$ mass) for such
a field is the one corresponding to the Compton wavelength of the order of the de Sitter radius $R$, giving {{{$m=m_{dS}=h/Rc$}}} which would
correspond to a quantum temperature $T_q=m_{dS} c^2 /4\pi^2 k\simeq m_{ds} R^2 H^2/k$. This is of the same order of magnitude of
the classical temperature $T_c= H^2 R^2 m_{dS}/2.26k$ calculated before, for classical particles of mass $m_{dS}$ in a box extending
to the cosmological horizon. This allows us to interpret the classical temperature $T_c=H^2 L^2 m/2.26k$ as a classical analogue,
in de Sitter spacetime, of the Hawking-Unruh effect.

We have derived the above analogy for a two dimensional de Sitter spacetime. However, the above considerations can be easily
extended to $dS_4$ provided the classical particles are taken to be rigid spheres of some small but nonzero radius.

In addition, the expression for $T_c$ has been derived under the assumption that the classical particles are non relativistic.
Strictly speaking, this approximation is not justified. Indeed, since $R\simeq 10^{28} cm$, the quantum de Sitter temperature
$T_q$ is of the order of $10^{-29}\ {}^o{\rm K}$, whereas the de Sitter mass $m_{dS}$ is of the order $10^{-65}g$. Therefore, the average
velocity of our de Sitter particles at the de Sitter temperature is comparable with the speed of light, so that they are highly relativistic.
This fact can be readily understood by noting that, as the walls of the box approach the cosmological horizon, their speed relative
to the comoving coordinates approaches $c$. Therefore, when the first of the de Sitter particle hits the wall it bounces back with
a highly relativistic speed, which is then transmitted to the other $dS$ particles through particle collisions and further collisions
with the walls themselves. Nonetheless, since we are only concerned with orders of magnitude, we still claim that our crude estimates relating
classical and quantum de Sitter temperatures are justified.

Finally note that our effective vacuum particles can in some sense be viewed upon as the lightest detectable particles in a de Sitter
background. Indeed, in such a background the largest uncertainty in position is $\Delta x\simeq R$, whereas $\Delta p\simeq mc$, so that
the Heisenberg principle gives $m\gtrsim h/Rc$. It is not clear what this exactly means, though one may boldly suggest that quantum
effects in de Sitter forbid the existence of lighter particles. More presumably, it may mean that the semiclassical description of
lightest particles is too na{\"\i}ve and that strong quantum effects do come into play.

\section{Collisions and decays.}\label{sec:collisions}
We consider the collision of two ingoing particles which gives rise to the production of  a certain number of outgoing particles
\begin{eqnarray}
b_1 + b_2\longrightarrow c_1+ c_2+\ldots +c_N \ .
\end{eqnarray}
The particles $b_i$, with masses $m_i$,  are described by
geodesic curves ending at the collision point $\X$, which is
also the starting point of the $N$ geodesics describing the
outgoing particles $c_f$ with masses  $\tilde m_f$. We assume
the collision point to be the common zero of the proper time of
all particles involved in the process, namely
$\X=X_i(0)=X_f(0)$. Denoting by $(\chi_i, \zeta_i)$ and by
$(\xi_f,\eta_f)$ the pairs of normalized null vectors
parameterizing the ingoing and outgoing particles we have
\begin{eqnarray} \zeta_i =\chi_i-\frac {2m_i}{k_i R}\X\ , \
i=1,2\ ; \qquad\ \eta_f =\xi_f -\frac {2\tilde m_f}{k_f R} \X\
,\ f=1,2,\ldots,N\ ,\label{32} \end{eqnarray} and the
quantities which are conserved along each geodesic are
\begin{eqnarray} K_i =\frac {k_i c}R \X\wedge \chi_i\ ,\ i=1,2\
;\qquad\ K_f=\frac {k_f c}R \X\wedge \xi_f\ ,\ f=1,2,\ldots,N\ .
\label{33} \end{eqnarray} Solving the collision problem amounts
to finding the outgoing vectors $\xi_f$ given the ingoing ones
$\chi_i$. At the collision point the total covariant
energy-momentum four-vector must be conserved:
\begin{eqnarray}
\pi_1^\mu + \pi^{\mu}_2 =\sum_{f=1}^M \pi^{\mu}_f. \label{totalmomentum}
\end{eqnarray}
Here we have introduced a local coordinate system $x^{\mu}$,
$\mu=0,1,2,3$, so that $\pi^{\mu}=m\frac {dx^{\mu}}{d\tau}$
(respectively, $\pi^{\mu}=kc\frac {dx^{\mu}}{d\sigma}$) for any
given massive particle of timelike (respectively, massless particle
of lightlike) worldline $x^{\mu}(\tau)$ (respectively
$x^{\mu}(\sigma)$) on $\dS$. In terms of the embedding in $\MMM$, at
the collision point $\bar X$ we have, for a given particle,
\begin{eqnarray} \left.K_{AB}\right|_{X=\X}=\frac 1R \left(X_{A}
\left.\frac {\partial X_{B}}{\partial x^{\mu}}-X_{B} \frac {\partial
X_{A}}{\partial x^{\mu}}\right)\right|_{x=\bar{x}} \pi^{\mu}\ ,
\label{Kp}
\end{eqnarray} where $X_A=X_A(x^\mu(\tau))$ or
$X_A=X_A(x^\mu(\sigma))$ depending on whether the particle is
massive or massless. By summing over all ingoing and outgoing
particles and using (\ref{totalmomentum}) we find the simple
relation
\begin{eqnarray}
K_1+K_2 = \sum_{f=1}^N K_f . \label{scattering}
\end{eqnarray}
Similarly, for the decay $b\longrightarrow c_1+c_2+\ldots+c_N$
of a single particle
\begin{eqnarray}
K=\sum_{f=1}^N K_f\ . \label{decay}
\end{eqnarray}
Note that $K_{AB}K^{AB}=-2m^2 c^2$. This relation  replaces in
$\dS$ the Minkowskian one $\pi_{\mu}\pi^{\mu} = m^2 c^2$. Then,
choosing the normalization constants $k_i$ and $k_f$ equal for
all particles, Eq.(\ref{33}) allows us to write the
conservation equations (\ref{scattering}) and (\ref{decay})
respectively as
\begin{eqnarray} (\chi_1+\chi_2 -\sum_{f=1}^N \xi_f )\wedge \X=0, \label{38}
\\ (\chi -\sum_{f=1}^N \xi_f )\wedge \X=0\ .
\label{39} \end{eqnarray} Though equations (\ref{scattering}) and
(\ref{38}) are equivalent to equation (\ref{totalmomentum}) they
have the advantage of being expressed in an intrinsic form. To
further clarify their meaning it is interesting to find the explicit
expressions of the null vectors $\chi_i$ and $\xi_f$ corresponding
to a particular choice of the collision event $\X$. For example,
choosing $\bar X=\oo$ equation (\ref{38}) becomes equivalent to
\begin{eqnarray}
\chi_1^\mu+\chi_2^\mu=\sum_{f=1}^N \xi_f^\mu \
, \ \mu=0,1,2,3 \ . \label{311}
\end{eqnarray}
From Eq. (\ref{32})
\begin{eqnarray} \zeta^\mu
=\chi^\mu\ \;\;\mu=0,1,2,3, \mbox{    and    }\
\zeta^4=\chi^4 -\frac {2m}k  \label{312}
\end{eqnarray} (we have omitted the index $i=1,2$ for notational simplicity). Since $\chi$ and $\zeta$ are null vectors,
if $m \neq 0$ this relation implies $\chi^4=-\zeta^4 =\frac
{m}k$. Therefore, we have
\begin{eqnarray} && \chi =\left(\chi^0, \vec \chi, \frac {m}k
\right)\ , \;\;\; \zeta=\left(\chi^0, \vec \chi, -\frac {m}k
\right)\ , \label{313}
\end{eqnarray} with
\begin{eqnarray}
(\chi^0)^2 -(\vec \chi )^2 =\frac {m^2}{k^2} \
. \label{314}
\end{eqnarray}
By using the parametrization (\ref{212}) it follows that
\begin{eqnarray} \left. m \frac {dX^\mu}{d\tau}\right|_{\tau=0}
=kc\chi^\mu =q^\mu , \qquad  \left. m\frac
{dX^4}{d\tau}\right|_{\tau=0}=0,\label{315}
\end{eqnarray} with
\begin{eqnarray} q^2=(q^0)^2 -(\vec q )^2 =m^2 c^2 \ .
\end{eqnarray}
In a small neighborhood of $\oo$ in $dS_4$ we choose local
coordinates $x^\mu$ defined by $x^\mu=X^\mu$,$\mu = 0,1,2,3$. Since
the plane $X^4=R$ is tangent to $dS_4$ at $\oo$ we have $\partial
\oo / \partial x^\mu|_{\oo}=0$ so that, at $\oo$, the metric of
$dS_4$, expressed in terms of the coordinates $x^\mu$, is given by
$ds^2|_{\oo}=(\eta_{AB}dX^A dX^B)|_{dS_4,\oo}=\eta_{\mu\nu}dx^\mu
dx^\nu$. Then the $x^\mu$ are locally Lorentzian at $\oo$ and
${dX^\mu}/{d\tau}|_{\tau=0}={dx^\mu}/{d\tau}|_{\tau=0}$ where
$x^\mu(\tau)$ is the parametrization of the geodesic at $\oo$.
Hence, equation (\ref{315}) tells us that $q^\mu$ can be interpreted
as the components (in the chosen frame)  of the Lorentzian
four-momentum at $\oo$ of the particle moving along
the geodesic $x^\mu (\tau)$. This interpretation applies to zero mass particles as well.\\
Then, denoting by $q^\mu_i$ and $\tilde q^\mu_f$ respectively
the four-momenta at the collision point $\X=\oo$ of the
incoming and outgoing particles relative to the coordinates
$x^\mu$ we have
\begin{eqnarray}
\chi_i=\frac 1{kc}(q^0_i,\ \vec q_i,\ m_i c )\ ,\ i=1,2, \label{317}
\end{eqnarray}
and
\begin{eqnarray}
\xi_f =\frac 1{kc}(\tilde q^0_f,\ \vec {\tilde q}_f,\ \tilde m_f c )\ ,\ f=1,2,\ldots,N \label{318}
\end{eqnarray}
and the conservation equation (\ref{311}) becomes
\begin{eqnarray}
q_1^\mu +q_2^\mu =\sum_{f=1}^N \tilde q^\mu_f \label{319}
\end{eqnarray}
expressing once more the equivalence of (\ref{38}) to
(\ref{totalmomentum}). Similar considerations apply to the
decay (\ref{decay}).

 The expressions of the incoming and
outgoing null vectors $\chi_i$ and $\xi_f$ in the general case,
when the collision point $\X$ is arbitrary, can be obtained by
applying to (\ref{317}) and (\ref{318}) an arbitrary
five-dimensional Lorentz transformation.

In conclusion, it is worthwhile noting that since any Lorentzian
manifold is locally inertial, at the classical level the
conservation laws in de Sitter point particle collisions express
nothing more than the usual total energy-momentum conservation in
the process, so that $\Lambda$ plays no role here. The situation
is drastically different in the quantum case due essentially to
the spread of wave packets. For example, in de Sitter particle
decay the decay amplitude depends on $\Lambda$ and the presence of
curvature allows in some cases for a non-zero probability for an
unstable particle of mass $m$ to decay into particles whose total
mass is larger than $m$, a process which is strictly forbidden in
Minkowski spacetime due to energy-momentum conservation
\cite{Bros:2006gs}.

Finally, as regards the geodesic motion of a single particle, it is
important to remark that the explicit expressions
\begin{eqnarray}
&&\xi = \frac{1}{kc}(q^0,\vec{q},mc), \nonumber \\
&&\eta = \frac{1}{kc}(q^0,\vec{q},-mc), \label{320}
\end{eqnarray}
of the components of the pair of normalized null vectors $\xi$ and
$\eta$ characterizing the particle geodesic $X(\tau)$ when $\XX$ is
chosen at the origin (\ref{origin}) as well as their corresponding
expressions for arbitrary $\XX$, which are obtained by applying to
(\ref{320}) a suitable five-dimensional Lorentz transformation,
depend solely on the choice of $\XX$ and do not make reference to
any particular local coordinate system on $dS_4$. Instead the
introduction of one such suitable system about $\XX$ is made
necessary for the correct physical interpretation of the components
of $\xi$ and $\eta$.

\subsection{Detection} In a collision process any outgoing
particle is not detected, and its properties measured, at a
collision point $\X$. Instead, the detection takes place at
some other event  far away from $\X$. In particular, if we
measure the energy and the momentum of the particle, we need a
formula which relates these quantities at the point of
measurement to the same quantities at the production point. To
avoid being monotonous we illustrate the procedure with a
lively example. Consider the $p  p$ scattering
$$
p + p \longrightarrow p + p +a+b+c \ ,
$$
and suppose that we are searching for an intermediate process
\begin{eqnarray}
p+ p \longrightarrow p+ p +Z  \longrightarrow p + p+a+b+c \
,\label{ppx}
\end{eqnarray}
where $Z$ is a massive particle decaying into the triple $a,b,c$
with a very short lifetime, so that it cannot be directly detected.
Then, by (\ref{decay})
$$
K_Z =K_a+K_b+K_c\ ,
$$
so that
$$
2K^2_Z:=(K_{aAB}+K_{bAB}+K_{cAB})(K_a^{AB}+K_b^{AB}+K_c^{AB})=-2m_Z^2
c^2
$$
must hold. Assume we look at a large number of such processes and
that we are able to measure experimentally $K_{aAB}, K_{bAB}$ and
$K_{cAB}$ in each individual process. Then, plotting the number
density of processes  $dn/dK$ as a function of the invariant mass
$Q_Z:=\sqrt {-K^2_Z}$, we should find a resonance at $Q_Z=m_Z c$. As
a well-known example of a reaction of the type (\ref{ppx}) we may
mention the process
$$
p+ p \longrightarrow p+ p +Z  \longrightarrow p + p+\pi^+ +\pi^- +
\pi^0\ ,
$$
where $Z$ can either be one of the mesons $\eta(547)$ or
$\omega(782)$ or some broader resonance.
See e.g. ref.\cite{Barberis:1998in}.\\
The experimental problem of measuring the quantities $K_a$,
$K_b$, $K_c$ could be tackled as follows. To fix ideas,
consider just one particle, which we suppose to detect at an
event whose local coordinates are $x^{\mu}_1$. Barring
intrinsic indeterminacies the detection measures the position
$x_1^{\mu}$ and the momentum $\pi_1^{\mu}(\tau_1)$. Then
$K_{AB}$ is determined by (\ref{Kp}) as
\begin{eqnarray}
K_{AB}= \left. K_{AB}\right|_{x=x_1} = \left. \frac 1R
\left(X_{A} \frac {\partial X_{B}}{\partial x^{\mu}}-X_{B}
\frac {\partial X_{A}}{\partial x^{\mu}}\right)\right|_{x=x_1} \pi^{\mu} (\tau_1)\ .
\label{Kp1}
\end{eqnarray}
This formula can  be used to relate the covariant momentum at
the point of measurement to the one at the collision point.
Indeed, if $x^\mu_0$ are the local coordinates of the collision
event and $\pi^{\mu} (0)$ the covariant momentum of the
particle at the same point, then
\begin{equation}
\left. K_{AB}\right|_{x=x_1}=
\left. \frac 1R \left(X_{A} \frac {\partial X_{B}}{\partial x^{\mu}}-X_{B}
\frac {\partial X_{A}}{\partial x^{\mu}}\right)\right|_{x=x_0} \pi^{\mu} (0)\ .
\end{equation}
By multiplying both sides of this equation by
$$
\left. \frac 1R \left(X^{A} \frac {\partial X^{B}}{\partial x^{\nu}}-X^{B}
\frac {\partial X^A}{\partial x^{\nu}}\right)\right|_{x=x_0},
$$
and summing over $A$ and $B$ we find
\begin{eqnarray}
\pi^{\mu}(0)=G^{\mu}_{\nu}(x_0,x_1) \pi^{\nu} (\tau_1) \ .\label{transfer}
\end{eqnarray}
Here
$$
G^{\mu}_{\nu}(x_0,x_1)=-\frac {1}{2R^2} g^{\mu\rho}(x_0) \left.
\left(X^{A} \frac {\partial X^{B}}{\partial x^{\rho}}-X^{B} \frac
{\partial X^{A}}{\partial x^{\rho}}\right)\right|_{x=x_0} \left.
\left(X_{A} \frac {\partial X_{B}}{\partial x^{\nu}}-X_{B} \frac
{\partial X_{A}}{\partial x^{\nu}}\right)\right|_{x=x_1}\ ,
\label{gmunu}
$$
where
$$
g_{\mu\nu}(x)=\eta_{AB}\frac {\partial X^{A}}{\partial x^{\mu}}\frac {\partial X^{B}}{\partial x^{\nu}}\ ,
$$
is the metric on $dS_4$ in the given coordinates.
Formulas (\ref{Kp}), (\ref{scattering}), (\ref{decay}) and (\ref{transfer}) hold for lightlike particles as well.\\
As an example, choose the local coordinates $x^{\mu}=\{ct,x^i\}$
to be the flat ones:
\begin{eqnarray}
X(t,x^i)=\left\{\begin{array}{lll}
X^0&=&  R\sinh \frac {ct}R +\frac {\vec x^2}{2R} e^{\frac {ct}R} \ ,\\
X^i &=&  e^{\frac {ct}R} x^i \ ,  \\
X^4 &=&  R\cosh \frac {ct}R -\frac {\vec x^2}{2R} e^{\frac {ct}R}
\ .
\end{array}\right.
\label{flat-coor}
\end{eqnarray}
If for simplicity we restrict ourselves to the two dimensional
case in flat coordinates and choose $x_0=(0,0)$ and $x_1=(ct,x)$
we find
\begin{equation}
G^{\mu}_{\nu} (x_0,x_1)=\left(
\begin{array}{cc}
1 & e^{2\frac {ct}R} \frac xR \\
\frac xR & e^{\frac {ct}R} \cosh \frac {ct}R +e^{2\frac {ct}R} \frac {x^2}{2R^2}
\end{array}
\right)\ . \label{propagator}
\end{equation}
In particular, consider the case of a photon transmitted from $x_0$  to $x_1$.
To express the momenta in terms of inertial frames at rest in each
point with respect to the given local coordinates,
we introduce the zweibein $e^0=cdt$, $e^1=e^{\frac {ct}R}dx$. The inertial
energy-momentum $\hat \pi^{\mu}$ has components $\hat \pi^0=\pi^0$ and $\hat \pi^1=e^{\frac {ct}R}\pi^1$. In particular, in $x_0$,
$\hat \pi^\mu (0)=\pi^\mu (0)$ and
\begin{eqnarray}
&& \hat \pi^0(0)= \hat \pi^0 + e^{\frac {ct}R} \frac xR \hat \pi^1\ , \label{prima}\\
&& \hat \pi^1(0)= \frac xR \hat \pi^0 +\left(\cosh \frac {ct}R
+e^{\frac {ct}R} \frac {x^2}{2R^2}\right)\hat \pi^1\
.\label{seconda}
\end{eqnarray}
Obviously $x_1$ cannot be any point, but must lie on a lightlike
geodesic starting from $x_0$. It can be easily found putting
$$
\hat \pi^0(0)=\hat \pi^1(0) =\frac {h\nu_0}c \ , \qquad \hat
\pi^0=\hat \pi^1 =\frac {h\nu}c \ ,
$$
in ($\ref{prima}$) and ($\ref{seconda}$) and solving for $x=x(t)$. This gives
$$
x(t)=R(1-e^{-\frac {ct}R})\ .
$$
Using this in (\ref{prima}) we finally obtain
\begin{eqnarray}
\nu=e^{-\frac {ct}R} \nu_0 \ .
\end{eqnarray}
This is the redshift measured by the observer at $x_1$: the photon
emitted with frequency $\nu_0$ at $x_0$ is perceived as a photon
of frequency $\nu$ by the observer at $x_1$.

\newpage
\section{Energy.}\label{sec:energy}

In Einstein's special relativity the energy of a particle is defined
(and measured) relative to an arbitrary given Lorentz frame, it
being the zero component of a four-vector. In physical terms, a
Lorentz frame can be seen as  an ideal global network of (free)
particles relatively at rest and carrying  clocks that stay forever
synchronized. This picture does not extend to the de Sitter case
where frames are defined only locally.

However, the maximal symmetry of the de Sitter universe  allows for
the energy of a pointlike particle to be defined relative to just
one reference massive free particle understood conventionally to be
at rest (the sharply localized observer). Below we will compare this
definition with the ones obtained in various coordinate patches by a
more standard Lagrangian approach.

The procedure amounts to fixing arbitrarily a timelike reference
geodesic (the geodesic of the particle ``at rest''). Let us denote
by $u$ and $v$ the future oriented null vectors which identify such
geodesic; the energy of the free particle (\ref{parametrizzazione})
with respect to the reference geodesic is defined as follows:
\begin{eqnarray}
E=E_{(\xi,\eta)}(u,v)=-\frac {c\,K_{(\xi,\eta)}(u,v)}{u\cdot v} \ .\label{energymass}
\end{eqnarray}
We have that $E_{(u,v)}(\xi,\eta)=E_{(\xi,\eta)}(u,v)$ which can be
interpreted as the symmetry between the active and passive point of
view. In particular, the proper energy is
$E_{(\xi,\eta)}(\xi,\eta)=mc^2 $, as it should be. To further
elaborate this definition let us choose an origin $\YY = Y(0) $ on
the reference geodesic and denote by $\lambda$ the scalar such that
\begin{equation}
u\cdot v  = 2\lambda^2,\;\;\;v  =  u\, -\frac{2\lambda\YY}{R}.\label{455}
\end{equation}
As before, fixing $\lambda$ removes the scale arbitrariness in
the choice of $u$ and $v$ and it follows that
\begin{equation}
Y(\tau) = R\frac { u\, e^{\frac {c\tau}R}-v \,e^{-\frac
{c\tau}R}}{2\lambda}= \YY e^{-\frac{c\tau}{R}}\ + \  \frac {
Ru}{\lambda}  \sinh \frac{c \tau}{R}
 \label{ref-geod}
\end{equation}
Proper times in (\ref{1fondamentale}) and (\ref{ref-geod}) are
of course not to be confused. Taking into account Eqs.
(\ref{215}) and  (\ref{455}) it follows that

\begin{eqnarray}
E = -mc^2 \frac{(\xi \wedge \eta )(u,v)}{(\xi\cdot \eta)( u\cdot v)
}
=-  {\frac { kc^2}{\lambda R^2}\, (\xi\wedge  \XX)}(u,\YY).
\end{eqnarray}
Finally, by inserting into this expression Eq. (\ref{imply})
and the analogous relation
\begin{equation}
\lambda = -\frac{1}{R}(u\cdot \YY )
\end{equation}
we get the expression
\begin{equation}
E =  {m c^2 }\frac{( u \cdot \XX)(\xi \cdot \YY )-(\XX\cdot \YY
)(\xi \cdot u)}{(\xi\cdot \XX)(u\cdot \YY )}, \label{energymass1}
\end{equation}
which is an alternative form of (\ref{energymass}).\\ \vskip 10 pt
We now wish to refer the energy $E$ of the particle to a given
coordinate patch. This can be done as follows. Suppose a local frame
$(t,x^i)$ has been selected so that the embedding of $\dS$ in $\MMM$
is given by $X^A (P)=X^A (t,\vec x)$; to fix ideas let us perform
this choice so that the event $t = 0, \vec x = 0$ is the "origin"
$\oo$ of the de Sitter manifold.

Then, we define the energy $E$ of a particle relative to the given
frame as the energy of the particle w.r.t. the particle (observer)
at rest at the origin, i.e.  w.r.t. the reference geodesic passing
through the origin with zero velocity ($\vec x(0)= 0$, $\frac {d\vec
x}{dt} (0)= 0$). We work out a few explicit examples.

\subsection{Flat coordinates.}

The flat coordinate system $\{t,x^i\}$ is defined by
(\ref{flat-coor}).
In these coordinates  the de Sitter geometry is
that of a flat exponentially expanding Friedmann universe:
\begin{eqnarray}
ds^2= c^2 dt^2 -e^{ {2ct}/R} \delta_{ij} dx^i dx^j = c^2 dt^2 -a^2(t) \delta_{ij} dx^i dx^j .
\end{eqnarray}
The reference geodesic (\ref{ref-geod}) through the origin
$Y(t=0,x^i = 0)=(0,0,0,0,R)$ with zero velocity is uniquely
associated to the choices $\YY=(0,0,0,0,R)$ and
$u=\lambda(1,0,0,0,1)$. With such a choice for $\YY$ and $u $, Eq.
(\ref{energymass1}) is explicitly written as follows:
\begin{equation}
E = \frac{kc^2}{R}(\xi^0\XX^4-\xi^4 \XX^0). \label{energy4}
\end{equation}
Noting that
\begin{equation}
\xi = \frac{m}{kR}\left(\XX +
\frac{R}{c}\frac{dX(0)}{d\tau}\right). \label{xi-via-X}
\end{equation}
and using Eq. (\ref{flat-coor}) we readily find
\begin{eqnarray}
&& E=
{mc^2}\frac {dt}{d\tau}
-\frac c R x^i p_i=
\frac {mc^2}{\sqrt {1-a^2(t) \frac {v^2}{c^2}}}
-\frac c R x^i p_i \label{flatenergy}
\end{eqnarray}
where we have set
\begin{eqnarray}
 v^i=\frac {dx^i}{dt}\, ,
 \qquad p_i =-m{e^{2 {ct}/R}}\frac {dx^i}{d\tau}=-\frac {m{a^2(t)}v^i}{\sqrt {1-a^2(t) \frac {v^2}{c^2}}} \ .
 \label{momenta}
\end{eqnarray}
In Section \ref{sec:momentum} we will show that
\begin{eqnarray}
p^i=-\frac{1}{a^2(t)}p_i =
\frac{mv^i}{\sqrt{1-a^2(t)\frac{v^2}{c^2}}} \label{momenta1}
\end{eqnarray}
can be interpreted as the de Sitter version of the linear momentum
(in flat coordinates). In the limit $R\longrightarrow \infty$,
(\ref{flatenergy}) and (\ref{momenta1}) go over into the usual
Minkowskian expressions of the energy and momentum.

That (\ref{flatenergy}) can be interpreted as the correct de
Sitter energy of the particle is confirmed by noting that it is
the conserved quantity associated to the invariance of the
particle action (\ref{freeaction}) under time translation.
Indeed, since in flat coordinates the spatial distances dilate
in the course of time by the exponential factor
$e^{\frac{ct}{R}}$, the expression of an infinitesimal symmetry
under time evolution is
\begin{eqnarray}
&&t \longrightarrow t + \epsilon, \nonumber \\
&&x^i \longrightarrow x^i - \frac{c}{R}x^i\epsilon.
\label{time-trans}
\end{eqnarray}
The action
\begin{eqnarray}
S=-mc \int \sqrt {1- e^{\frac{2ct}R} \frac {v^i v^j}{c^2}
\delta_{ij}}\ dt \label{freeaction1}
\end{eqnarray}
is invariant under (\ref{time-trans}) and, by Noether's theorem,
the corresponding constant quantity is precisely
(\ref{flatenergy}).

For a massless particle, using Eq. (\ref{massless-limit}) we find
\begin{eqnarray}
&& E=kc^3 \frac {dt}{d\sigma}-\frac c R x^i p_i \ ,
\label{energy-null}
\end{eqnarray}
where we have set
\begin{equation}
 p_i =-kc{e^{2\frac {ct}R}} ({dx^i}/{d\sigma})\ .
\label{momentum-null}
\end{equation}
 In particular,
note that $ \frac {d\vec x}{dt} \cdot \frac {d\vec x}{dt}=c^2
e^{-2\frac {ct}{R}} \ . $ To find the relation between $t$ and
$\sigma$, we take the derivative with respect to $\sigma$ of the
relation defining the cosmic time $ X^0+X^4=Re^{\frac {ct}{R}}\ ,
$ and use (\ref{nullgeod}) to obtain
\begin{eqnarray}
\frac {dt}{d\sigma}=\frac 1c e^{-\frac {ct}R} (\xi^0+\xi^4)\ .
\end{eqnarray}
Inserting this into the expression of the energy we find
\begin{equation}
 E=kc^2 (\xi^0+\xi^4) e^{-\frac {ct}R}-\frac c R x^i p_i , \ \ \ \
\makebox{} \qquad
p_i =-k(\xi^0+\xi^4){e^{\frac {ct}R}}\frac {dx^i}{d t}\ .
\end{equation}
In the flat limit $R\longrightarrow \infty$ we have $E\longrightarrow kc^2 (\xi^0+\xi^4)$ so that,
if we associate a frequency to the de Sitter massless particle, we have
$h\nu=kc^2 (\xi^0+\xi^4)$ and finally
\begin{equation}
E=h\nu e^{-\frac {ct}R}-\frac c R x^i p_i\ , \ \ \ \
p_i =-\frac {h\nu}{c^2} {e^{\frac {ct}R}}\frac {dx^i}{d t}\ .\label{nullmomenta}
\end{equation}
In the limit $R\longrightarrow \infty$ we obtain the usual Minkowskian expressions
\begin{eqnarray}
E=h\nu\ , \qquad\ \vec p=\frac {h\nu}c \vec n\ .
\end{eqnarray}

\subsection{Spherical coordinates.}
Let $\{t, \omega^\alpha \}$, $\alpha=1,\ldots,4$, be such that
\begin{eqnarray}
\left\{\begin{array}{lll}
X^0 &=& R\sinh \frac {ct}R \ ,\\
 X^\alpha&=&  R\cosh \frac {ct}R \, \omega^\alpha \ ,
 \end{array}\right.
\end{eqnarray}
where $\omega^\alpha \omega^\beta \delta_{\alpha\beta} =1$,
that is the $\omega^\alpha$ is a vector on the sphere $S^3$ of
unit radius. Concretely
\begin{eqnarray}
\left\{\begin{array}{lll}
\omega^1&=&\sin \chi^1 \sin \chi^2 \cos \chi^3\\
\omega^2&=&\sin \chi^1 \sin \chi^2 \sin \chi^3 \ ,\\
\omega^3&=&\sin \chi^1 \cos \chi^2 \ ,\\
\omega^4 &=&\cos \chi^1\
\end{array}\right.
\end{eqnarray}
This coordinate system covers the whole de Sitter manifold. The
geometry is that of a closed Friedmann universe undergoing an
epoch of exponential contraction which is followed by an epoch of
exponential expansion:
\begin{equation}
ds^2=c^2 dt^2 -R^2 \cosh^2 \frac {ct}R \left\{ \left(d{\chi^1}\right)^2
+\sin^2 \chi^1 \left[\left(d{\chi^2}\right)^2 +\sin^2 \chi^2 \left(d{\chi^3}\right)^2\right]\right\}.
\end{equation}

The initial point and the lightlike vector identifying the reference
geodesic are once more $\YY =\oo$ and $u=\lambda(1,0,0,0,1)$, and
therefore the energy is again given by Eq. (\ref{energy4}).
Expressing $\xi$ in terms of $X$ and $dX/d\tau$ it follows that
\begin{eqnarray}
&& E=mc^2 \frac {\omega^4-\frac R{2c} v^4 \sinh \frac {2ct}R}{\sqrt
{1-\frac {R^2}{c^2} \cosh^2 \frac {ct}R v^\alpha v^\beta \delta_{\alpha\beta}}}\ ,
\label{ballenergy}
\end{eqnarray}
where $\alpha,\beta=1,2,3,4$ and $v^\alpha=d\omega^\alpha/dt$.\\
Again, expression (\ref{ballenergy}) can be  recovered as the
conserved quantity associated to a time translation plus a
rescaling of the $\omega$'s that together leave invariant the
action
\begin{eqnarray}
S=-mc^2 \int \sqrt{1 -\frac {R^2}{c^2} \cosh^2 \frac {ct}R \Omega_{ij} w^i w^j}\ dt \ ,
\end{eqnarray}
where $d\Omega^2 = \Omega_{ij} d\chi^i d\chi^j$ and  $w^i
=\frac {d\chi^i}{dt}$.

\subsection{Static (black hole) coordinates.}
This is the coordinate system originally introduced by de Sitter in
his 1917 paper \cite{desitter}. It describes a portion of the de
Sitter manifold as follows:
\begin{eqnarray}
\left\{\begin{array}{lll}
X^0 &=& R \sqrt{1-\frac {r^2}{R^2} } \sinh \frac {ct}R \ ,\\
X^i&=& r^i \ , \quad i=1,2,3\ , \\
X^4&=& R \sqrt{1-\frac {r^2}{R^2} } \cosh \frac {ct}R \ ,
\end{array}\right.\end{eqnarray}
where $r^2=\sum_{i=1}^3 r^i r^i$. With these coordinates the metric
exhibits a bifurcate Killing horizon at $r=R$:
\begin{eqnarray}
ds^2 =\left( 1-\frac {r^2}{R^2}\right) c^2 dt^2 -\frac {dr^2}{\left( 1-\frac {r^2}{R^2}\right)}
-r^2 (d\theta^2 +\sin^2 \theta d\phi^2) \ .
\end{eqnarray}
We choose the same origin and reference geodesic as before and
find
\begin{eqnarray}
E=mc^2 \left( 1-\frac {r^2}{R^2}\right) \frac 1{\sqrt{1-\frac
{r^2}{R^2}-\frac {(\vec{r}\cdot\dot{\vec{r}})^2}{(R^2-r^2)c^2}
-\frac {\dot {\vec r} \cdot \dot {\vec r}}{c^2} }}\
.\label{blackenergy}
\end{eqnarray}
In these coordinates, the action for a massive free particle is
\begin{eqnarray}
S=-mc^2 \int \sqrt{1-\frac {r^2}{R^2}-\frac
{(\vec{r}\cdot\dot{\vec{r}})^2}{(R^2-r^2)c^2} -\frac {\dot {\vec
r} \cdot \dot {\vec r}}{c^2}   }\ dt\ .
\end{eqnarray}
Here the dot means derivation with respect to $t$.
This action is invariant under time translations and the associated conserved energy
coincides with (\ref{blackenergy}).\\
We leave it as an exercise to find the analogue of expressions (\ref{ballenergy}) and
(\ref{blackenergy}) for massless particles. As
mentioned in the introduction, a fourth example can be found in
\cite{Gursey:1964} where the expression of the energy is given
in terms of stereographic coordinates.

\section{A possible definition of momentum.}\label{sec:momentum}
Whereas, as shown in section \ref{sec:energy}, the energy of a
particle can be defined relative to an arbitrary fixed reference
geodesic, and therefore in a frame independent manner, no similar
characterization can be given for the linear momentum. Instead, a
definition of momentum for a de Sitter particle necessarily requires
the selection of some coordinate system. In Minkowskian relativity
energy and momentum are defined as the conserved quantities
associated with the invariance of the action respectively under
infinitesimal inertial time and space translations. In addition, in
the de Sitter case, for a class of reference frames which in the
limit $R\longrightarrow \infty$ become inertial, we have seen that
the particle energy arises again as the conserved quantity
associated to time translations (depending on the choice of the
particular coordinate system, the latter may or may not act on the
space coordinates as well). Therefore, it is natural to attempt to
define the de Sitter momentum in any such frame as the conserved
vector quantity which is associated with infinitesimal space
translations. This requires fixing the origin of the coordinate
system, since, due to the presence of curvature, changing the origin
affects the definition of space translations. Of course, in order
for such definition of momentum to be consistent, one must make sure
to recover the usual Minkowskian momentum in the limit
$R\longrightarrow \infty$. Then we search for Lorentz
transformations in the embedding spacetime $M(1,4)$ which generate
spatial translations of the origin, once the latter has been
identified. Specifically, let $(t,x^i)$ be local coordinates, $X^A
(t,x^i)$ the embedding functions and $O\equiv X^A(0)$ the origin. We
consider the submanifold defined by $t=0$. It defines an
hypersurface of $\dS$ which identifies an osculating hyperplane in
$O$. Any infinitesimal Lorentz transformation which leaves the
osculating plane invariant defines an infinitesimal translation of
$O$ which is transverse to the reference geodesic defining the
energy. As stated above we use such transformations to define the
momentum. We illustrate again the construction for the coordinate
systems of section \ref{sec:energy}.

\subsection{Flat coordinates.}
In this case, the $t=0$ surface defines the osculating plane $X^0+X^4=R$.
The Lorentz transformations leaving this hyperplane invariant are
generated by the infinitesimal rotations of the form $v\wedge v_i$,
where $v=(1,0,0,0,-1)$ and $v_i=(0,\vec e_i, 0)$, where $\vec e_i$ is
the standard basis of $\bR^{3}$. Then the momentum is
\begin{equation}
p_i:=K(v,v_i)\ ,
\end{equation}
which expressed in coordinates gives
\begin{equation}
p_i=-m{e^{2\frac {ct}R}}\frac {dx^i}{d\tau}\ .
\end{equation}
This expression coincides with (\ref{momenta}) and corresponds to the conserved
quantities associated to the invariance of the action under
spatial translations $x^i\mapsto x^i+a^i$. For a massless particle
we should use $\tilde K$ in place of $K$, finding the second of
formulas (\ref{nullmomenta}) as conserved momentum.
Note that we have a scale ambiguity in choosing the vectors $v_i,v$. However such ambiguity can be
fixed by requiring to obtain the usual Minkowskian expression in the $R\longrightarrow\infty$ limit.

\subsection{Spherical coordinates.}
In spherical coordinates the $t=0$ slice correspond to $X^0=0$.
We have $v=(0,0,0,0,1)$ and we can choose $v_i$ as before.
Then
\begin{equation}
K(v,v_i)=mc \frac {\xi_4 \eta_i-\eta_4\xi_i}{ \xi\cdot\eta} \ ,
\label{formula}
\end{equation}
and, in term of the coordinates,
\begin{equation}
p_i =mR\cosh^2 \frac {ct}R \left( \omega_4 \frac {d\omega_i}{d\tau}-\omega_i \frac {d\omega_4}{d\tau} \right)\ .
\end{equation}
The flat limit leaving the origin invariant can be easily
performed and it gives the correct momentum for Minkowski
spacetime.

\subsection{Black hole coordinates.}
Again the $t=0$ slice defines the osculating hyperplane $X^0=0$ and
$K(v,v_i)$ is once more given by (\ref{formula}). Then
\begin{eqnarray}
\!\!\!\!\!\!\!\!\!\!\!\!\!\!\!\!\!\!\!\!\!\!\!\!\!\!\!\!\!\!\!\!\!\!\!\!\!
p_i=\!m\!\left( \sqrt {1-\frac {r^2}{R^2}}  \frac {dr_i}{d\tau}
\cosh \frac {ct}R -\sqrt {1-\frac {r^2}{R^2}}  \frac {r_i}R c \sinh
\frac {ct}R -\frac 1{\sqrt {1-\frac {r^2}{R^2}}} \frac {r_i}{R^2}
\sum_{k=1}^3 r_k \frac {dr^k}{d\tau} \cosh \frac {ct}R \right)\ .
\cr
\end{eqnarray}


\section{Conclusions.}\label{sec:conclusions}
In this paper we have taken the stance that in the absence of
gravitation the spacetime arena in which all physical phenomena take
place is the (maximally symmetric) four dimensional de Sitter
manifold $dS_4$. Then, barring discrete spacetime operations such as
space reflection and time reversal, which are not exact symmetries
of nature, the corresponding relativity group is the connected
component of $SO(1,4)$, the (ten dimensional) de Sitter group. This
attitude springs naturally from very basic properties of all natural
phenomena, which so far have never been experimentally contradicted.
To wit, experiments and observations in the whole realms of physics,
astrophysics and cosmology point to the fact that local laws
governing natural phenomena do not depend on time and on the
location in space (spacetime homogeneity) and that no direction in
space is privileged compared to any other (isotropy of space). In
addition, all experiments also point to the local validity of the
principle of inertia, which states that no operational meaning
whatsoever can be given to a notion of absolute rest (invariance
under boosts). Therefore, the laws of physics should possess
a ten parameter invariance group acting on the four dimensional
spacetime continuum, which embodies all above symmetries. Working at
the Lie algebra level it has been shown by Bacry and L\'evy-Leblond
in a remarkable paper published almost 40 years ago
\cite{Bacry:1968zf} that such a group is uniquely determined up to
two undetermined parameters, a velocity scale (the invariant speed
$c$) and a length scale (the cosmological constant $\Lambda$). The
actual values of these parameters in nature are of course not fixed
by the basic symmetries and must be found experimentally. And,
indeed, though limiting values of $c$ and $\Lambda$ (such as $c=0,\
\infty$ and/or $\Lambda=0,\ \infty$) cannot be excluded a priori, it
comes as no surprise that the values of $c$ and $\Lambda$ determined
by the observations are well defined and finite. It would be
surprising if it were otherwise! It is a different (and to some
extent metaphysical) question why $c$ and  $\Lambda$ have the values
they have and not others. However we are not concerned with this
problem here.

It is then clear that, if  Minkowski space should be replaced by de
Sitter space and, correspondingly, the Poincar\'e group by the de
Sitter group, one is naturally led to a reformulation of the theory
of special relativity on these grounds
\cite{Aldrovandi:2004hd,Aldrovandi:2006vr,Guo:2003qm,Guo:2004pj,Guo:2006pa}.
However, compared to the Minkowski case, this task presents certain
complications which are essentially connected with the fact that in
the de Sitter case there exists no class of privileged reference
frames as are the Minkowskian inertial ones. Indeed, the associated
coordinate systems of any such hypothetical class of equivalent
frames should respect the basic symmetries of the spacetime
manifold. In particular, the homogeneity requirement would imply the
coordinate transformation between any two such equivalent frames to
be affine \cite{LeLe}, and this is impossible if the underlying
manifold is curved. Therefore, the absence of a privileged class of
equivalent frames suggests that, in de Sitter relativity, it would
be desirable, whenever possible, to characterize significant
physical quantities in an intrinsic way, namely in a manner
independent of the choice of any particular coordinate patch. In
this paper we have accomplished this for any set of independent
conserved quantities along the geodesic motion of a free de Sitter
particle and for the overall conservation of the total constants of
the motion in any particle collision. In particular, we have also
been able to give an intrinsic definition of energy of a de Sitter
particle, as the energy of any such particle relative to an
arbitrary selected reference particle chosen conventionally to be at
rest. In this respect, it is important to stress that in the same
way as there is a unique Lorentzian (i.e. relativistic)
generalization of the kinetic energy of a Galilean (i.e.
nonrelativistic ) particle, the de Sitter energy is the unique de
Sitter generalization of the Lorentzian energy, which arises from
the appearance of an intrinsic residual curvature of the spacetime
manifold.

We remark that, due to the smallness of  $\Lambda$, the actual
corrections to Einstein's special relativity which are brought about
by the presence of the cosmological constant, such as for instance
those embodied in formula (\ref{transfer}), are utterly tiny at
scales of laboratory experiments performed on earth or in space.
Specifically, formulas (\ref{prima}) and (\ref{seconda}) tell us
that to first order in $1/R$ the fourmomentum of a particle
traveling a distance $x$ is altered by a relative amount of the
order $x/R$. In particular, for example, since $R  \simeq 10 ^{28}
{\rm cm}$, the order of magnitude of the fractional frequency shift
of a photon traveling a distance $x$ in the de Sitter universe is
\begin{equation}
\frac{\Delta \nu}{\nu} \simeq 10^{-25}\  \     {\rm for} \  \  x
\simeq 10 {\rm m}
\label{87}
\end{equation}
and
\begin{equation}
\frac{\Delta \nu}{\nu} \simeq 10^{-16}\  \   {\rm for}\  \  x
\simeq 10 {\rm million\  km}\ .
\label{88}
\end{equation}
The figure $\Delta \pi/\pi  \simeq 10^{-25}$  would be relevant also
for particles produced in a collision in a particle accelerator such
as the Tevatron or LHC since the distance between the collision
point and the detector is typically of the order of a several meters. By
comparison, we recall that in the classical experiment by Pound and
Rebka  \cite{Pound:1960zz,Pound:1964zz,snider} devised to measure the
gravitational shift of a photon falling in the earth's gravitational
field  we have
\begin{equation}
\frac{\Delta \nu}{\nu} \simeq 10^{-15}
 \label{89}
\end{equation}
whereas in experiments with atomic clocks which monitor the
variation of the gravitational potential of the sun at the location
of the earth between perihelion and aphelion \cite{Ashby:2007} we
have
\begin{equation}
\frac{\Delta \nu}{\nu} \simeq 10^{-10}
 \label{90}\ .
\end{equation}
The precision with which the value of $\Delta \nu/\nu$ has been
measured in the Pound and Rebka experiment is of the order of one
percent, whereas values of $\Delta \nu/\nu$  in the range of the
figure of formula (\ref{90}) in experiments with atomic clocks as
are mentioned above are now tested with a precision of the order of
almost one part per million. This shows that, whereas there is no
chance to realistically test formula (\ref{transfer}) for $\Lambda$
in the gravitational field of the earth, even for falls of thousands
of kilometers, comparison of the ticks of atomic clocks set in
suitable eccentric orbits around the sun may in principle be able to
reveal an effect due to $\Lambda$ in a not too unforeseeable future.
Indeed, in such a hypothetical case, due to the known periodicities
it should be possible to filter out the effects from all other contributions,
gravitational and not.

Finally, we remark that while at the present epoch the effects due to $\Lambda$
are tangible only at cosmological scales, they might have been essential, even at
microscopic distances, during the period of inflationary expansion in the very
early universe, when the effective de Sitter radius was extremely small
($\simeq 10^{-27}-10^{-28} cm$), at which time, however, quantum effects are expected
to have been dominant \cite{Linde:2005ht}.

\vskip 1.5 cm
\subsection*{Acknowledgments}
We are indebted to F.~Hehl who stimulated our interest in the subject.

\newpage

\section*{Appendix A
} We show in this appendix by constructing an explicit example that
the components of the future oriented lightlike fivevectors $\xi$
and $\eta$ which identify a given timelike geodesic are fixed by the
assignement of the values of six independent parameters once $\xi$
and $\eta$ have been separately normalized according to equation
(\ref{211}) and the selection of the point of the geodesic
corresponding to zero proper time. Precisely, let $x^\mu=\{ct,r^i\}$
be a set of local coordinates on $\dS$ and consider a timelike
geodesic passing through $\vec r_0$ at time $t=0$ with velocity
$\vec v_0=d\vec r /dt|_{t=0}$. Parameterizing the geodesic as in
(\ref{parametrizzazione}) and choosing a suitable normalization for
$\xi$ and $\eta$, we can express this vectors in terms of the
initial conditions $\vec r_0$ and $\vec v_0$. Indeed
\begin{eqnarray}
& X(x^\mu(t)) =R \frac {\xi e^{\frac {c\tau(t)}R}-\eta e^{\frac
{-c\tau(t)}R}}{\sqrt {2\xi\cdot \eta}}\ . \end{eqnarray} Then,
setting $t=0$ and assuming $\tau(0)=0$, we find
\begin{eqnarray}
&& X (x_0) =\frac R{\sqrt {2\xi\cdot \eta}} (\xi -\eta)
\end{eqnarray}
and, taking the derivatives with respect to $t$ at $t=0$ we get
\begin{eqnarray}
&& \left. \frac {dX}{dt}\right|_{t=0} = \partial_\mu X (x_0) {\dot
x}^\mu_0 =\frac c{\sqrt {2\xi\cdot \eta}} (\xi +\eta) \dot \tau_0 \
.
\end{eqnarray} Choosing $k=m$ in (\ref{211}), and solving with
respect to $\xi$ and $\eta$ we find
\begin{eqnarray}
&& \xi =\frac {\partial_\mu X (x_0) v^\mu_0}{c\dot \tau_0} +\frac
{X(x_0)}R \ ,\cr && \eta =\frac {\partial_\mu X (x_0) v^\mu_0}{c\dot
\tau_0} -\frac {X(x_0)}R \ , \label{form}\end{eqnarray} where
$v^0=c$, $v^i= \dot {x}^i$ and \begin{eqnarray} c\dot \tau_0 =\left.
\sqrt { \frac {ds^2}{dt^2}}\right|_{t=0}\ .
\end{eqnarray}
The condition $\xi \cdot \eta=2$ and the choice of the initial point
$X(x_0)$ on the geodesic fixes the separate normalizations of $\xi$
and $\eta$ and, as a consequence, formulas (\ref{form}) express
$\xi$ and $\eta$ as functions of $\vec r_0$ and $\vec v_0$.

For example, if we choose static coordinates we have explicitly
\begin{eqnarray}
&& \xi^i =\frac 1{c\dot \tau_0} {v_0^i}+ \frac {r_0^i}{R} \ ,\\
&& \eta^i =\frac 1{c\dot \tau_0} {v_0^i}- \frac {r_0^i}{R}\ , \quad\
i=1,2,3
\end{eqnarray}
and similar formulas for $\xi^0$, $\eta^0$, $\xi^4$ and $\eta^4$,
where
\begin{eqnarray} \dot \tau_0=\left. \frac
{d\tau}{dt}\right|_{t=0}= \sqrt{1-\frac {r_0^2}{R^2}-\frac
{v_0^2}{c^2}-\frac {(\vec r_0 \cdot \vec v_0)^2}{c^2(R^2-r_0^2)}}=
\frac {mc^2}{E}\left(1-\frac {r_0^2}{R^2}\right)\ .
\end{eqnarray}
Note in particular that $ds^2$ is positive along timelike
curves so that setting \begin{eqnarray} \vec \rho=\frac 1R \vec
r\ , \qquad\ \vec \beta =\frac 1c \vec v\ , \end{eqnarray} we
have
\begin{eqnarray} 1-\rho^2 -\beta^2- \frac {\rho^2 \beta^2 \cos^2
\theta}{1-\rho^2}>0\ , \end{eqnarray} where $\theta$ is the
angle between $\vec r$ and $\vec v$ (and $\vec \rho$, $\beta$
and $\theta$ are functions of $t$). Then \begin{eqnarray} \beta
< \frac {1-\rho^2}{\sqrt {1-\rho^2 \sin^2 \theta}}\ .
\end{eqnarray} In particular for lightlike geodesics we find
\begin{eqnarray} \beta = \frac {1-\rho^2}{\sqrt {1-\rho^2
\sin^2 \theta}}\ , \end{eqnarray} so that massless particles
have velocities \begin{eqnarray} {1-\rho^2}\leq \beta \leq
\sqrt {1-\rho^2}\ , \end{eqnarray} depending on the angle
between the velocity and the position vector.
\section*{Appendix B}
In one dimension the de Sitter line element in flat coordinates is given by
\begin{equation}
ds^2=c^2 dt^2 -e^{\frac {2ct}R} dx^2=c^2 dt^2 -a^2(t) dx^2 \ .
\end{equation}
Then, denoting by $y$ the spatial coordinate with respect to which the rigid box has fixed length $2L$, we have
\begin{equation}
y= xe^{\frac {ct}R}
\end{equation}
so that the j-th particle of our classical comoving gas drifts towards one of the walls of the box according to the equation
\begin{equation}
y_j(t)=y_{0j} e^{\frac {ct}R}\ , \qquad\ y_{0j}=L\frac jN\ , \quad j=0,\pm 1,\pm 2,\ldots, \pm (N-1)\ .
\end{equation}
Hence the $\pm (N-1)-th$ particle, which is the one initially closest to the right (left) wall, reaches the latter (and bounces back
elastically) at the time
\begin{equation}
t_{N-1}=\frac Rc \log \frac {N}{N-1}
\end{equation}
thereby colliding after a while with the $\pm(N-2)-th$ particle, thus starting the thermalization process through further collisions
with the other particles. Now, the equation of motion of particle j is
\begin{equation}
\ddot y_{j}-\frac {c^2}{R^2} y_j=0
\end{equation}
corresponding to a potential
\begin{equation}
V(y)=-\frac 12 \frac {c^2}{R^2} y^2\ .
\end{equation}
Therefore, after some transient time the gas virializes itself, eventually reaching a thermal state at a temperature $T$ given by
\begin{equation}
\frac 12 kT =\langle K \rangle=m |\langle V \rangle|=m \frac 1{\int_{-L}^L \rho(y)dy}
\int_{-L}^L |V(y)|\rho(y) dy\ ,
\end{equation}
where $K$ is the kinetic energy and $\rho(y)$ is the mass density of the gas at equilibrium.
Since our gas is an ideal one its equation of state is
\begin{equation}
m p(y)=\rho(y) kT\ .\label{state}
\end{equation}
Then, eliminating the pressure from (\ref{state}) and the Euler equation
\begin{equation}
-\frac {dV(y)}{dy}=\frac 1{\rho(y)} \frac {d p(y)}{dy}
\end{equation}
we get
\begin{equation}
-\frac {dV(y)}{dy}=\frac {kT}m \frac {d}{dy} \log \rho(y)
\end{equation}
from which
\begin{equation}
\frac {kT}m \log \frac {\rho(y)}{\rho(o)} =|V(y)|=\frac {c^2}{2R^2}y^2\ .
\end{equation}
Hence
\begin{equation}
\rho(y)=\rho(0) e^{\frac {mc^2}{2kTR^2} y^2}
\end{equation}
so that
\begin{equation}
kT =\frac {mc^2}{R^2}\frac {\int_{-L}^L dy y^2 e^{\frac {mc^2}{2kTR^2} y^2}}{\int_{-L}^L dy e^{\frac {mc^2}{2kTR^2} y^2}}
\end{equation}
which can be written as
\begin{equation}
\int_0^{\frac LR \sqrt {\frac {mc^2}{2kT}}} dw w^2 e^{w^2} =\frac 12
\int_0^{\frac LR \sqrt {\frac {mc^2}{2kT}}} dw e^{w^2}\ .
\end{equation}
The solution of this equation is
\begin{equation}
\frac LR \sqrt {\frac {mc^2}{2kT}}=1.063
\end{equation}
namely{{{
\begin{equation}
T\simeq \frac {mc^2 L^2}{2R^2 k (1.063)^2} \simeq \frac {mH^2 L^2}{2.26 k}\ .
\end{equation}
}}}

\newpage
\bibliographystyle{ieeetr}
\bibliography{biblio2}
\end{document}